\documentclass[useAMS,usenatbib]{mnras}

\usepackage{graphicx}
\usepackage{amssymb}
\usepackage{amsmath}
\usepackage{datetime}
\usepackage[normalem]{ulem}
\usepackage{times}

\usepackage{color}
\usepackage{soul}
\definecolor{stan}{rgb}{0,0,1}

\definecolor{dylan}{rgb}{0.431,0.106,0.537}

\definecolor{asif}{rgb}{1.0,0.2,0}

\date{\currenttime \today}

\begin{document}

\title[Magnetic disruption of circumstellar discs]
{
Disruption of circumstellar  discs by large-scale stellar magnetic fields
}

 \author[A. ud-Doula et al.]
{
Asif ud-Doula$^{1,4}$\thanks{Email: asif@psu.edu},
Stanley P. Owocki$^2$, and
Nathaniel Dylan Kee$^3$
\\
\\
 $^1$ Penn State Scranton, 120 Ridge View Drive, Dunmore, PA 18512, USA.\\
 $^2$ Bartol Research Institute, Department of Physics and Astronomy, 
 University of Delaware, Newark, DE 19716, USA\\
 $^3$ Institut f\"ur Astronomie und Astrophysik, Eberhard Karls Universit\"at T\"ubingen, D-72076 T\"ubingen, Germany \\
 $^4$ LESIA, Observatoire de Paris, 5, place Jules Janssen, F-92195 Meudon, France
}

\def\<<{{\ll}}
\def\>>{{\gg}}
\def\wig{{\sim}}
\def\spose#1{\hbox to 0pt{#1\hss}}
\def\ltwig{\mathrel{\spose{\lower 3pt\hbox{$\mathchar"218$}}
     R_{\rm A}ise 2.0pt\hbox{$\mathchar"13C$}}}
\def\gtwig{\mathrel{\spose{\lower 3pt\hbox{$\mathchar"218$}}
     R_{\rm A}ise 2.0pt\hbox{$\mathchar"13E$}}}
\def\+/-{{\pm}}
\def\=={{\equiv}}
\def\mubar{{\bar \mu}}
\def\mustar{\mu_{\ast}}
\def\Lambar{{\bar \Lambda}}
\def\Rstar{R_{\ast}}
\def\Mstar{M_{\ast}}
\def\Lstar{L_{\ast}}
\def\Tstar{T_{\ast}}
\def\gstar{g_{\ast}}
\def\vth{v_{th}}
\def\grad{g_{rad}}
\def\glines{g_{lines}}
\def\Mdot{\dot M}
\def\mdot{\dot m}
\def\yr{{\rm yr}}
\def\ksec{{\rm ksec}}
\def\kms{{\rm km/s}}
\def\qad{\dot q_{ad}}
\def\qlines{\dot q_{lines}}
\def\solar{\odot}
\def\Msun{M_{\solar}}
\def\msbyr{\Msun/\yr}
\def\Rsun{R_{\solar}}
\def\Lsun{L_{\solar}}
\def\Be{{\rm Be}}
\def\Rpole{R_{p}}
\def\Req{R_{eq}}
\def\Rmin{R_{min}}
\def\Rmax{R_{max}}
\def\Rstag{R_{stag}}
\def\vinf{V_\infty}
\def\Vrot{V_{rot}}
\def\Vcrit{V_{\rm crit}}
\def\half{{1 \over 2}}
\newcommand{\beq}{\begin{equation}}
\newcommand{\eeq}{\end{equation}}
\newcommand{\beqa}{\begin{eqnarray}}
\newcommand{\eeqa}{\end{eqnarray}}
\def\phip{{\phi'}}

\maketitle

\begin{abstract}
Spectropolarimetric surveys reveal that 8-10\% of OBA stars harbor large-scale magnetic fields, but thus far no such fields have been detected in any classical Be stars.
Motivated by this, we present here MHD simulations for how a pre-existing Keplerian disc 
-- like that inferred to form from decretion of material from rapidly rotating Be stars --
can be disrupted by a rotation-aligned stellar dipole field.
For characteristic stellar and disc parameters of a near-critically rotating B2e star, we find that a polar surface field strength of just 10 G can significantly disrupt the disc, while a field of 100 G, near the observational upper limit inferred for most Be stars, completely destroys the disc over just a few days.
Our parameter study shows that the efficacy of this magnetic disruption of a disc scales with the characteristic plasma beta {(defined as  the ratio between thermal and magnetic pressure)} in the disc, but is surprisingly insensitive to other variations, e.g. in stellar rotation speed, or the mass loss rate of the star's radiatively driven  wind.
The disc disruption seen here for even a modest field strength suggests that the presumed formation of such Be discs by decretion of material from the star would likely be strongly inhibited by such fields; this provides an attractive explanation for why no large-scale fields are detected from such Be stars.
\end{abstract}


\section{Introduction}

Modern spectropolarimetric surveys have revealed  that 8-10\% of massive, early-type (OBA) main sequence stars harbor large-scale (often significantly dipolar) surface fields ranging from about $100$ to 
$10,000$\,G, with the incidence remarkably constant over a large range of stellar parameters
\citep[see, e.g.][]{WadNei16,MorCas15}.
Classical Be stars are a noteworthy exception to this, as no large-scale field has ever been unambiguously detected in this class, despite a sample of roughly $100$ closely surveyed stars \citep{WadPet16,NeiGru12}.
The present paper uses magnetohydrodynamical (MHD) simulations to explore the level of dipole stellar field strength needed to disrupt a Keplerian disc with properties inferred for such Be stars. 

A defining characteristic of Be stars is the double-peak emission in Balmer lines arising from a circumstellar decretion disc. This is thought to form from ejection of material \citep{RivBaa03,KeeOwo16b} 
from an underlying star that is rotating at a substantial fraction \citep[generally $> 70$\%;][]{TowOwo04}
of critical rotation, defined here by the speed $V_{\rm crit} = \sqrt{GM/R}$ for material to be in orbit near the equatorial surface radius $R$. 
There is substantial observational evidence \citep[e.g.][and references therein]{RivCar13} 
to confirm the theoretical expectation that 
discs around classical Be stars 
must follow a Keplerian form for orbital speed, $V_{\rm orb} (r) = V_{\rm crit} \sqrt{R/r}$.
The shear from the associated, radially declining orbital frequency $\Omega (r) = V_{\rm orb}/r = (V_{\rm crit}/R) (R/r)^{3/2}$  is inherently {\em incompatible} with the constant, rigid-body form $\Omega_{rot} = V_{rot}/R$ that a strong surface magnetic field would impose on ionized plasma near the stellar surface \citep{TowOwo05,udDOwo08}.

In the case of {\em accretion} discs (e.g. from protostars, or binary mass exchange), the strong radial decline in field energy, which scales as $B^2 \sim r^{-6}$ for a dipole field (and even more steeply for higher-order fields), means that such {\em external}\footnote{Small-scale, internal fields generated from the magneto-rotational instability (MRI) likely play a key role in the effective viscous transport of angular momentum that controls accretion.} magnetic fields can be largely ignored for modeling the inward viscous diffusion from the accretion mass source;
thus only in the near-surface regions can the underlying stellar magnetic field substantially influence the final accretion onto the surface, leading for example to bi-polar jets and/or broader wind outflows \citep[see, e.g.][]{ShuNaj94,RomUst09}.

But for Be discs that are thought to arise from the {\em decretion} of material ejections from the rapidly rotating underlying star, a sufficiently strong field can effectively choke off the formation of any disc before it reaches into a stable, Keplerian orbit.
This provides a potential rationale for explaining why the large-scale fields inferred for OBA stars are inherently inconsistent with the classical Be phenomenon.

Here we do not attempt to model this complex, generally highly variable decretion process.
Instead, the MHD simulations described in \S 2  explore how the introduction of a rotation-aligned dipole field affects a pre-existing Keplerian disc, assuming standard stellar and disc parameters expected for a B2e star. (See \S 2.1 and Table 1.).
A key result  (\S 2.2) is that a dipole field with polar strength $B=100$\,G at the stellar surface completely destroys the disc, while a field of $B=1$\,G has little effect.
The intermediate, transitional, `critical' case of $B_{\rm crit}=10$\,G,  disrupts the {\em inner} disc, but still leaves some material in Keplerian orbit away from stellar base.
Our parameter studies (\S\S 2.3-2.4) show surprisingly little effect from varying various parameters, e.g.:
equatorial rotation speed $V_{\rm rot}$ (\S 2.4.1);
the mass loss rate ${\dot M}$ from the star's radiatively driven stellar wind (\S 2.4.2); 
imposing a split monopole vs. dipole field from the stellar surface (\S 2.4.3);
and reducing numerical grid resolution in latitude to augment numerical reconnection (\S 2.4.4).
But  increasing disc {\em density} or {\em temperature} requires then a larger field to disrupt the disc, with $B_{\rm crit}^2 \sim \rho$ or T (\S 2.3).
This indicates that the {\em plasma $\beta$ }($\equiv P_{\rm gas}/P_{\rm mag} \sim \rho T/B^2$) is a key controlling dimensionless parameter, with the transitional case of $B_{\rm crit} =10$\,G having $\beta \approx 20$ for disc base gas pressure $P_{\rm gas}$ just above the surface ($r \gtrsim R$) in the equatorial disc mid-plane. 
We conclude  (\S 3) with a brief discussion of the potential implications of our study for magnetic disruption of other kinds of Keplerian discs (e.g. formed from protostellar accretion or binary mass exchange), along with an outlook for future work.

\begin{table}
\centering
\begin{tabular}{|| l l l ||}
\hline \hline
luminosity & L & 5000 L$_\odot$  \\
mass & M & 9 M$_\odot$  \\
stellar radius & R & 5 R$_\odot$  \\
temperature & T & 20,000 K\\
isothermal sound speed & c$_s$& 16.5 km s$^{-1}$ \\
initial peak disc density & $\rho_o$& $ 10^{-11.1}$ g cm$^{-3}$\\
near-star orbital speed & $V_{\rm crit}$ & 585 km s$^{-1}$ \\
rotation speed at equator & $V_{\rm rot}$ & $0.9 V_{\rm crit}$ \\
wind mass loss rate & $\dot{M}_{}$ & $1.4 \times 10^{-10}$ M$_\odot$ yr$^{-1}$  \\
[0.5ex]
\hline
\end{tabular}
\caption{Stellar and disc parameters for standard B2e star model.}
\label{table:1}
\end{table}

\section{MHD Simulations of Disc Disruption}
\label{sec:MHDsims}

\subsection{Numerical specifications for standard B2e model}
\label{sec:stdmodel}

The MHD simulations here use the same basic code \citep[ZEUS-3D;][]{StoNor92} and methods employed in our previous studies of magnetic channeling of radiatively driven stellar winds from OB stars \citep{udDOwo02,udDOwo08,udDOwo09}.
Now, however, instead of simply introducing an initial dipole stellar magnetic field to an existing radial wind outflow, we  also include a pre-existing  Keplerian disc.

This is generated by first defining an analytic disc model with a characteristic peak density
$\rho_i$ ($= 2 \times 10^{-11}$\,g\,cm$^{-3}$ in our standard model) in the equatorial mid-plane at the disc base (just above the equatorial stellar surface), with then a power-law radial fall off as $\rho_{} \sim r^{-3.5}$, as expected for a viscous decretion disc
\citep[see][and references therein]{KeeOwo16a}.
The stratification in height $z$ above a mid-plane radius $r$ is set to a standard gaussian form $\rho(r,z) = \rho(r,0) \exp^{-z^2/2H^2}$, where the disc scale height is given by $H = r c_s/V_{\rm orb}$,
with the sound speed $c_s \equiv \sqrt{kT/\mu} \approx 16.5$\,km\,s$^{-1}$ for our standard isothermal model with fixed temperature $T=20,000$K, and a mean molecular weight $\mu$ about 0.6 of a proton mass.
Since the near-surface orbital speed $V_{\rm orb} = V_{\rm crit} \approx 585$km s$^{-1} \gg c_s$, the inner disc is geometrically quite thin, with $H/R \approx 0.028$, and thus an initial full opening angle of just about 3.2 degrees.
Because we assume a fixed temperature and sound speed, while the Keplerian orbital speed declines as $V_{\rm orb} (r) \sim r^{-1/2}$, the outer disc flares up as this opening angle increases with $\sqrt{r}$.

\begin{figure*}
\begin{center}
\includegraphics[scale=0.50]{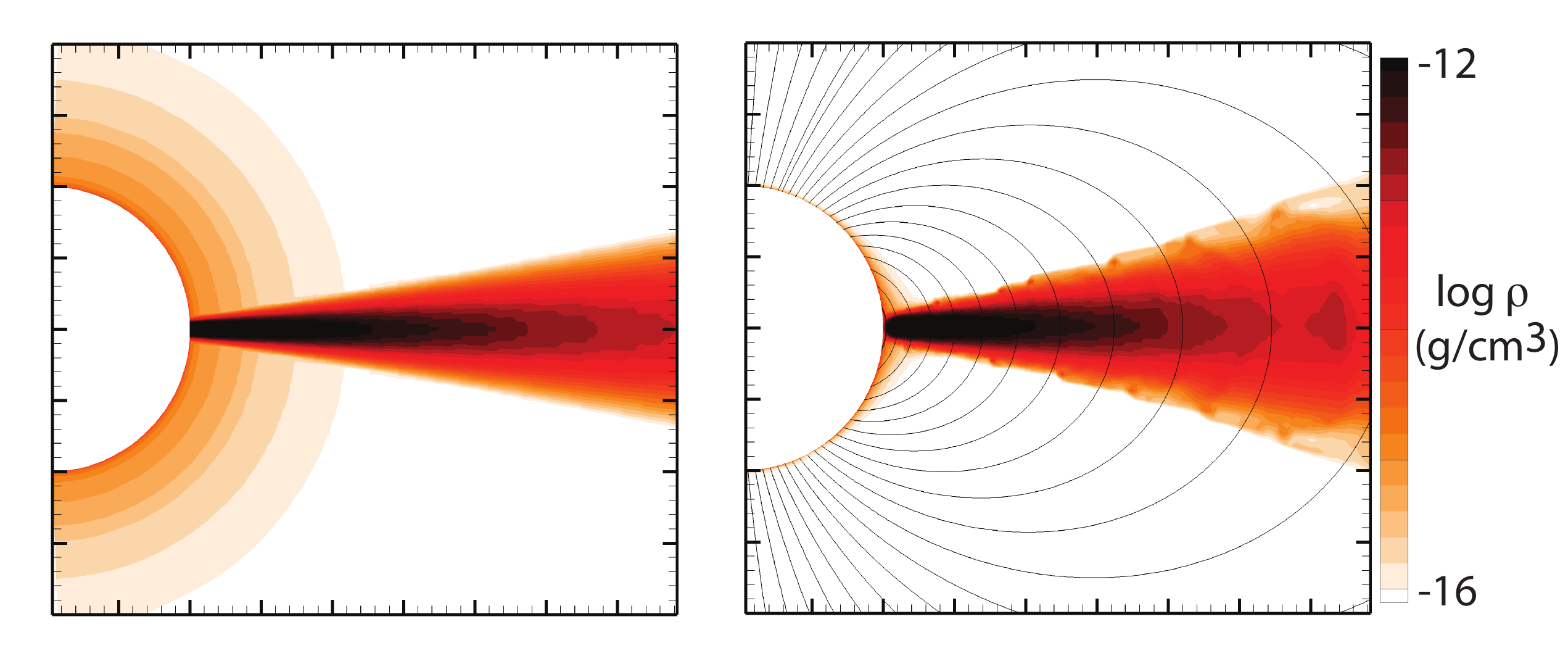}
\caption{Comparison of the mass density $\rho$ in the analytic model for disc+wind (left panel) with the hydrodynamically relaxed state (right panel) that includes the dynamical interaction between the disc and wind. This right panel now also includes the dipole field lines that are applied in the initial condition for the MHD simulations here.
}
\label{fig:rhoinits}
\end{center}
\end{figure*}

In our simulations, the surface of the disc also interacts with the radiatively driven stellar wind, leading to a gradual ablation of the disc over timescales of months or years 
\citep{KeeOwo16a}.
To minimize here any initial variations from this interaction, we first relax the model with pure hydrodynamical simulations for 1000 ks before introducing any magnetic field, which thus defines the initial time $t=0$ in our MHD simulations.
We additionally use only the radial component of line-acceleration, since \cite{KeeOwo16a} demonstrated that the non-radial components influence the ablation the most.
For the standard model, Figure \ref{fig:rhoinits} compares the analytic density distribution for the disc+wind with the hydrodynamically relaxed state used as an initial condition for the MHD simulations.
This hydrodynamically relaxed disc now has a characteristic peak density $\rho_o = 10^{-11.1}$\,g\,cm$^{-3}$, modestly reduced (by about a factor 0.4) from the value $\rho_i$ for the analytic initial model.

In this standard model, the underlying B2e star is assumed to have a rapid, near-critical rotation $V_{\rm rot} = 0.9 V_{\rm crit} = 527$\,km\,s$^{-1}$ (See Table 1.);
but to avoid ``staircasing'' effects on  wind initiation from an oblate stellar surface \citep{OwoCra94},
we ignore any rotationally induced stellar oblateness, and so simply set all boundary values for field and density at a spherical surface with a fixed stellar radius $R=5 \Rsun$.
For consistency, we also ignore any gravity darkening in the radiative driving of the stellar wind.
As in our previous simulations of wind-field interaction \citep{udDOwo02},
we use a generalized form of the line-driving formalism of 
\cite{CasAbb75} (hereafter CAK),
e.g. to account for integration over the finite stellar cone angle.
We choose the line opacity parameters ($k$ and $\alpha$ in the CAK notation, or ${\bar Q}$ and $\alpha$ in the notation developed by \cite{Gay95}),
so that the stellar luminosity, mass and radius lead to an overall mass loss rate ${\dot M} = 1.4 \times 10^{-10} \Msun$\,yr$^{-1}$, thought to be characteristic for our standard B2e star.

As in our previous wind-field simulation models, the lower boundary conditions fix the base density, but allow the radial velocity to vary (within an assumption of constant gradient) to adjust to the radiatively driven mass flux. The radial magnetic field is fixed to that of a base dipole, with the latitudinal and azimuthal field allowed to adjust, as needed, to the base wind outflow. 
The outer boundary conditions simply extrapolate quantities outward, both in the supersonic wind and stationary disc.

The 2D spherical mesh has 400 radial zones that increase in size by 1\% per zone, ranging from the stellar surface radius $R$ to an outer radius of $15 R$.
The latitudinal grid has 100 zones from pole to pole, with highest concentration at the equator, set to an equatorial step size of 0.01 radian (0.57 degree), which then increases by 4.5\% per zone to the poles.

\subsection{Results for standard models with B=1, 10, and 100 G}
\label{sec:stdresults}

\begin{figure*}
\begin{center}
\includegraphics[scale=0.650]{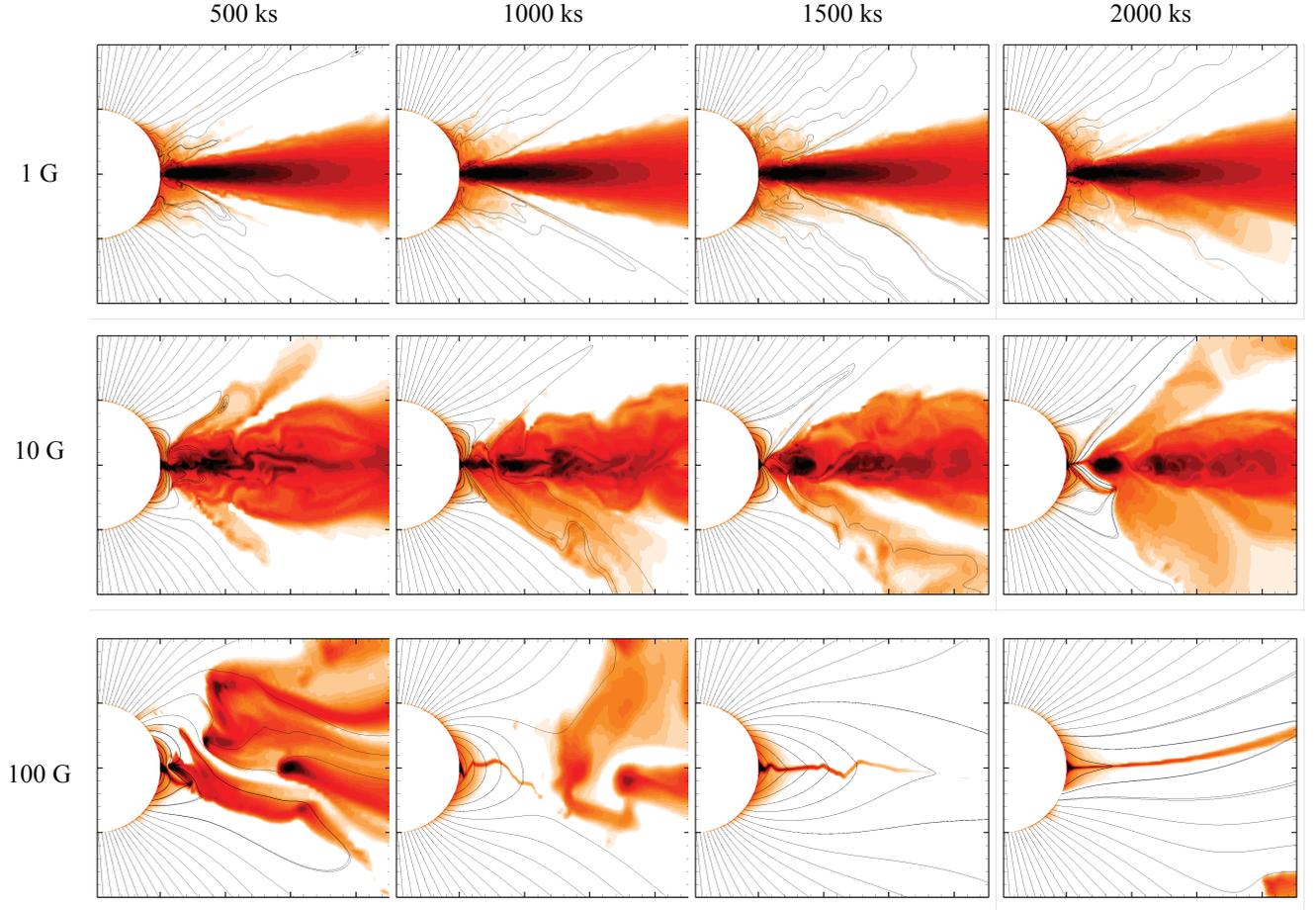}
\caption{Comparison of disc density for the standard B2e star model with polar field strengths of 1, 10 and 100  G (top, middle and bottom rows),  at time snapshots 500, 1000, 1500 and 2000 ks (left to right columns).
The density is rendered on a log scale over the same range as shown by the colorbar in figure \ref{fig:rhoinits}.
The lines denote magnetic field from the stellar surface.
}
\label{fig:rhovst}
\end{center}
\end{figure*}

Within the above specifications for the standard B2e parameters given in Table 1, 
Figure \ref{fig:rhovst} shows the spatial variation of density (over 4 dex on a log scale) at time snapshots $t=500$, 1000, 1500 and 2000 ks.
The upper, middle, and bottom rows compare results for models with polar surface field strengths of respectively $B=1$, 10, and 100 G.
Note that the lowest field of just 1\,G has only a minor effect on the disc, while the largest field of $100$\,G completely destroys it; indeed, the remaining equatorial density shown at large times ($t>1500$\,ks)  comes from the magnetically channeled, compressed wind, not from the initial Keplerian disc.
For the intermediate, transitional case with a critical field $B_{\rm crit} = 10$\,G, the inner disc is substantially disrupted, but there remains a significant residual density in the outer disc even at late times.

The disc evolution seen here is the result of a rather complex interaction between the disc and field, which along the disc surface also includes interaction with the outflowing wind.
The entrained field can augment the wind ablation that, even in pure hydrodynamical models, gradually erodes the disc over timescales of months or years \citep{KeeOwo16a}.
For stronger fields, this erosion become rapid enough to deplete the disc density over dynamical timescales, leading to feedback that further enhances the disruptive effects of the field, and so results in a complete destruction of the disc for the strong-field case.

The upshot thus seems to be that, for the chosen standard parameters for disc density and temperature, the fields in this range from 1-100 G represent a transition from little to full disruption.
Let us next examine further how this depends on various assumed parameters for the disc, as well as those associated with the stellar rotation, wind, field configuration, and even numerical grid resolution.

\subsection{Parameter study for disc mass evolution}
\label{sec:paramdmdr}

\begin{figure}
\includegraphics[scale=0.65]{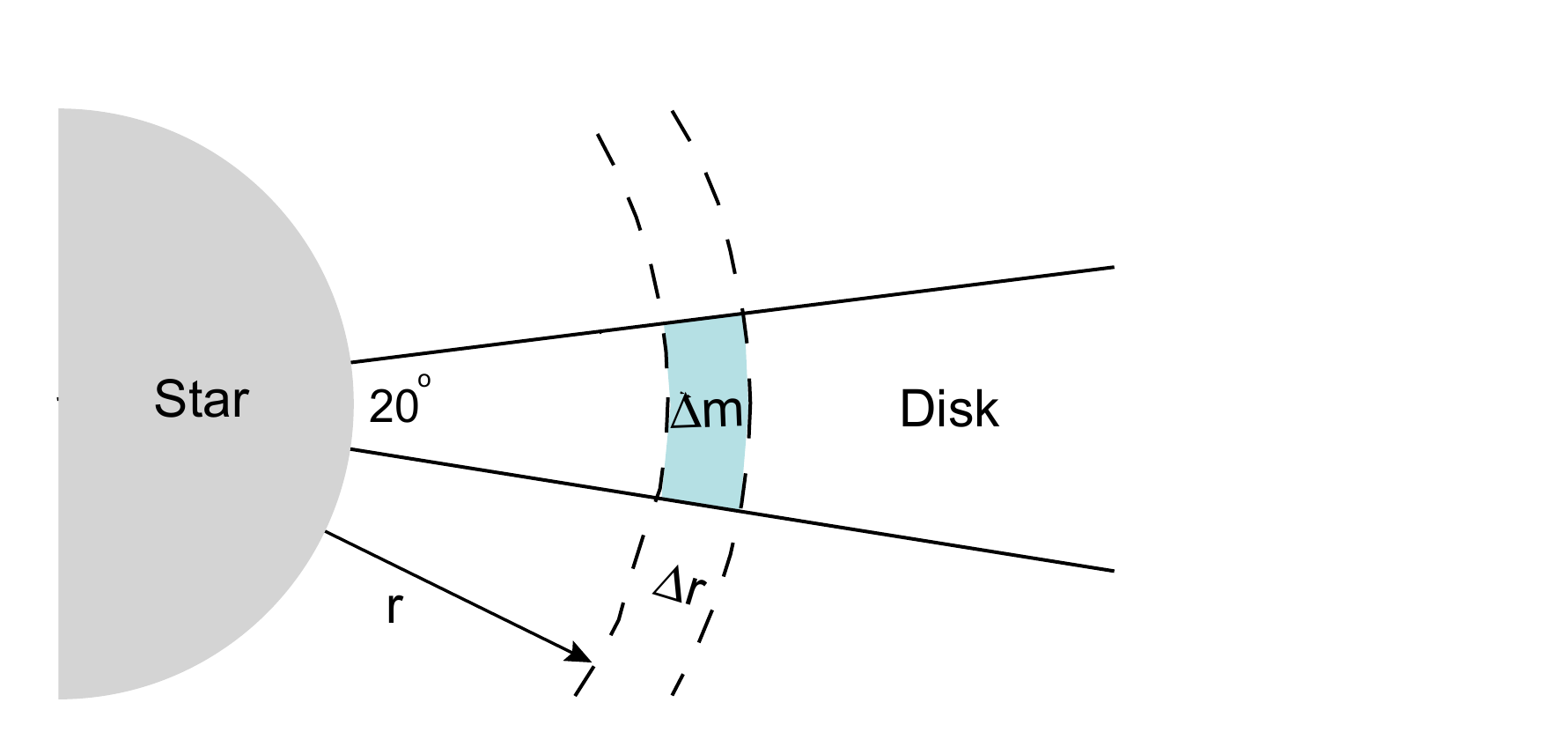}
\caption{
Schematic showing definition of equatorial mass distribution $dm/dr$ as the mass within an angle range of $20^o$ centered on the equatorial disc.
}
\label{fig:dmdr-def}
\end{figure}

For this broader parameter study, it is helpful to augment the direct plots of the time evolution of density at fixed time snapshots with a rendition of the full time and radius variation of the radial mass distribution $(dm/dr) (r,t)$ of the equatorial disc.
As in our previous analysis of magnetic wind channeling from a rotating star \citep{udDOwo08},
we define, for any arbitrary time $t$, the mass distribution about some latitudinal range (here taken to be $\theta = 90 \pm 10^o$) about the equator by
\beq
\frac{dm}{dr} (r,t) \equiv 2 \pi r^2 \int_{80^o}^{100^o} \, \rho(r,\theta,t) \sin \theta \, d\theta
\, .
\label{eq:dmdr-def}
\eeq
Figure \ref{fig:dmdr-def} gives a schematic drawing of this definition.

\begin{figure*}
\begin{center}
\vfill
\includegraphics[scale=0.85]{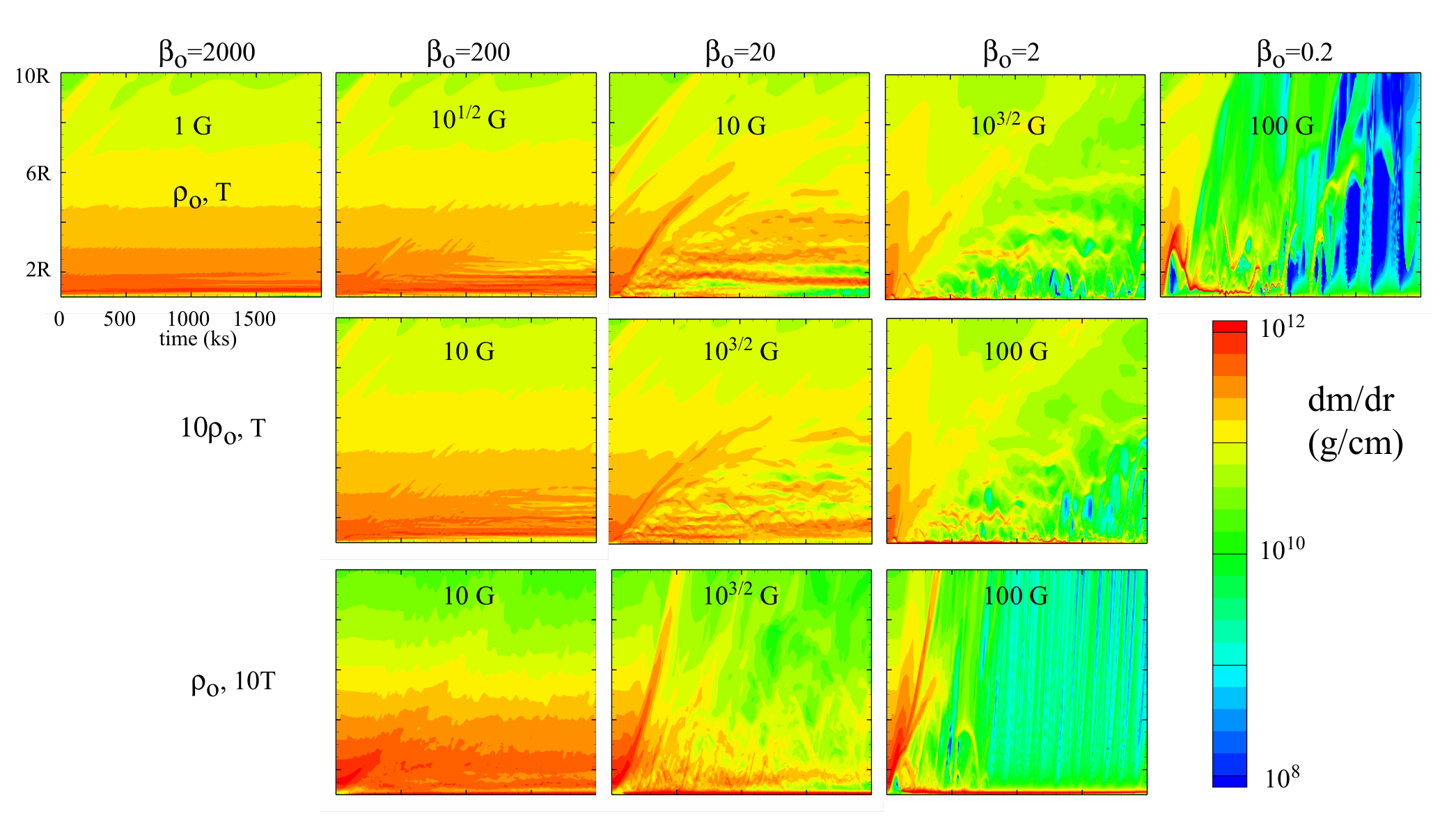}
\caption{
Radius and time evolution of equatorial mass distribution $dm/dr$, plotted on a log scale with 4 decades variation from red to blue.
The top row shows results for the standard B2e star with a disc base density 
$\rho_o = 10^{-11.1}$g\, cm$^{-3}$, 
labelled by values of  the magnetic field at the polar surface.
The top left panel shows the disc is largely unaffected by a 1~G field, 
while the top right panel shows that a 100~G field is sufficient to completely destroy the disc.
The middle panel shows the transitional case with a critical field of $B_{\rm crit} = 10~G$ that partially disrupts the disc, effectively removing material from the inner disc, but leaving some of the outer disc still intact.
The middle row shows models with 10 times higher base density $\rho_o$ (with colorbar rescaled accordingly), 
while the bottom row shows models with standard $\rho_o$ but 10 times higher temperature $T$.
Note that both changes require a half-decade increase in field strength to give a similar level of disc disruption as the standard case. 
This indicates that the plasma beta is a key controlling parameter for such magnetic disruption of the disc.
The top label gives the associated numerical values of $\beta_{\rm o} \equiv (\rho_o kT/\mu)/(B_{eq}^2/8\pi)$,
where $B_{eq}=B/2$ is the equatorial value of the magnetic field at the stellar surface.
}
\label{fig:dmdr-basic}
\end{center}
\end{figure*}

For the various models of our parameter study, Figure \ref{fig:dmdr-basic} then shows colorscale renditions of the variation of $dm/dr$ with radius and time over the model simulation range\footnote{Our simulations generally extend to 15 R, but to avoid occasional backeffects at the outer boundary, we limit analyses to a maximum radius of 10 R, well below any such outer boundary effects.} $r=1-10\,R$ and $t=0-2000$\,ks.
The upper row shows again the standard model cases, adding now intermediate cases with half-dex field increments $B=10^{1/2}$G and $B=10^{3/2}$G to the $B=1$, 10, and 100 G models discussed above.
The colorbar for $dm/dr$ varies logarithmically over 4 decades, with maximum value (red) at $dm/dr=10^{12}$\,g\,cm$^{-1}$ for the standard models cases, but a factor ten higher for models with ten times higher base density, as shown in the middle row.

The colorscales for each panel now show quite vividly the contrast  and progression between the weak-field limit with little  disc disruption ($B=1$G, top left panel) and the strong-field limit with complete disc destruction ($B=100$\,G, top right panel).
The intermediate case with critical field $B_{\rm crit} = 10$\,G shows that the near-surface disc is effectively disrupted, but there remains residual material in the outer disc.

As shown in the lower two rows, our study of parameter variations focuses within a half-dex variation about the critical value for partial disc disruption.
But now the disc is taken to have either a factor ten higher base density, $10 \rho_o$ (middle row),
or an artificially high, factor 10 hotter temperature, $10 T = 200,000$\,K (lowermost row).
Note then that the critical field needed for partial disruption is accordingly now a factor $\sqrt{10}$ higher.
 Thus models that are a half-dex lower and higher than this critical value now correspond to the weaker vs.\ stronger disc disruption case of the standard model in the top row.

The very top of each column is labeled with the associated, common characteristic value of the plasma beta, defined here as $\beta_{\rm o} \equiv (\rho_o kT/\mu)/(B_{eq}^2/8\pi)$,
where $B_{\rm eq}=B_{\rm pole}/2$ is the equatorial value of the magnetic field at the stellar surface.
The fact that the overall morphologies  within a column are quite similar indicates that the plasma beta represents a key controlling parameter for setting the efficacy of magnetic disc disruption.

It may at first seem surprising  that the onset, critical case for disruption occurs for a $\beta_{\rm o} = 20$ that is well above unity, with thus the equatorial magnetic pressure still much smaller than the initial gas pressure at the near-surface disc mid-plane. But of course this initial disc gas pressure is an overall {\em maximum} value, and so is not really representative of the disc at higher radii, or away from the equator, or indeed at later times after any initial disc erosion.
On the other hand, it seems quite appropriate that even for this somewhat biased definition, a plasma beta of order unity results in the total destruction of the disc, as shown by the uppermost right panel with $\beta_{\rm o} = 0.2$.

\begin{figure*}
\begin{center}
\includegraphics[scale=0.470]{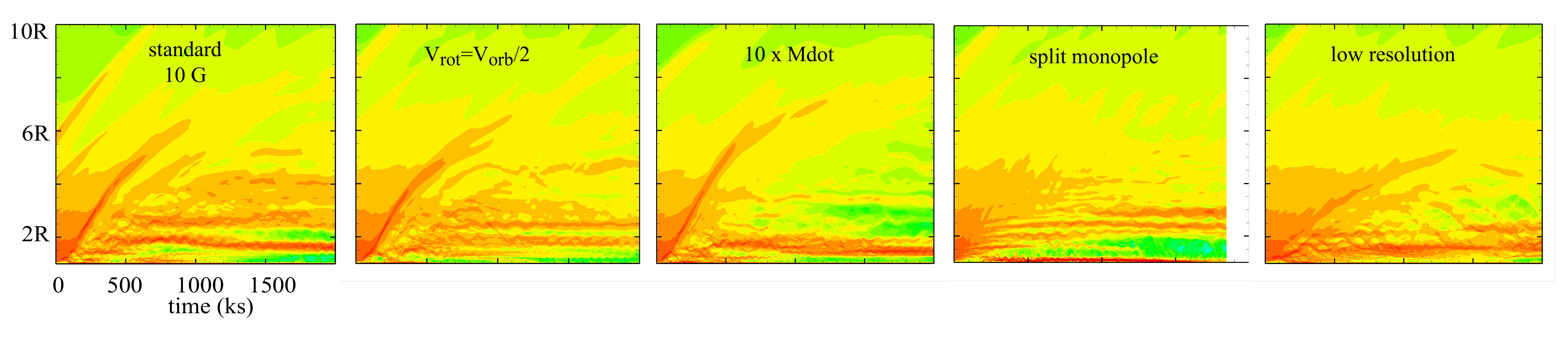}
\caption{
Radius and time evolution of equatorial mass distribution $dm/dr$ for the standard model case that has the critical field strength $B_{\rm crit}=10$ G.
The leftmost panel for the standard model is 
compared with various models with just one variation from this standard case.
Specifically, the panels progressively to the right show models with:
 lower rotation speed  $V_{\rm rot} = V_{\rm orb}/2$;
10 times higher wind mass loss rate;
a split monopole (vs.\ dipole) field;
and 1/2 the numerical grid resolution in latitude.
The similarities in the time and radius evolution of $dm/dr$ among all the panels indicate the magnetic field disruption of the disc is largely insensitive to these parameter variations.
}
\label{fig:dmdrvary}
\end{center}
\end{figure*}

\subsection{Relative insensitivity to other parameter variations}
\label{sec:nulleffects}

We have also explored the effects of varying other parameters that seem plausibly likely to be important for disc disruption, like stellar rotation speed, wind mass loss rate, field topology, and even numerical grid resolution.

The models shown in Figure \ref{fig:dmdrvary} all assume the same critical field value of $B_{\rm crit} =10$\,G, along with the set of stellar parameters for the standard model with just one specific variation.
The leftmost panel reproduces again the $dm/dr$ variation for this critical case of the standard model.

\subsubsection{Effect of reduced rotation speed}
\label{sec:lowrot}

The next panel to the right shows results for a case with a reduced stellar rotation speed, now set to $V_{\rm rot} = 0.5 V_{\rm crit}$ (instead of the 90\% critical rotation of the standard model).
Since the stellar field is rigidly locked to this  rotation, this implies a greater relative speed, $\Delta V = \Omega_{\rm rot} R - V_{\rm crit} \sqrt{R/r}$, between the rotationally swept stellar field and the orbiting material in the Keplerian disc.
For the standard case, this relative speed has a base value $\Delta V = 0.1 V_{\rm crit} = 58.5$\,km\,s$^{-1}$;
but in this slower rotation model this becomes five times higher, $\Delta V = 0.5 V_{\rm crit} = 292$\,km\,s$^{-1}$, 
implying factor 25 increase in the relative kinetic energy.
Nonetheless, the $dm/dr$ for this case shown in the second panel from the left appears surprisingly similar, indeed nearly indistinguishable, when compared against the leftmost panel representing the standard case with near-critical rotation.

\subsubsection{Effect of increased wind mass loss}
\label{sec:highmdot}

The middle panel of Figure \ref{fig:dmdrvary} shows a similar insensitivity of magnetic disc disruption to a factor ten increase in the wind mass loss rate.
In our previous simulations of the competition between a wind outflow and a stellar dipole field \citep{udDOwo02},
we identified a dimensionless `wind magnetic confinement' parameter $\eta_\ast \equiv B_{eq}^2 R^2/{\dot M} v_\infty$ (with $v_\infty$ the terminal wind speed in the case of no magnetic field).
This sets a characteristic Alfv\'en radius $R_A \approx \eta_\ast^{1/4} R$ for maximum closed magnetic loops, with the wind above $R_A$ stretching the dipole field lines open into a radial `split monopole' configuration.
Since this reduces the magnetic flux through the equatorial regions of the disc, one might expect that raising or lowering $\eta_\ast$ (and thus $R_A$) would enhance or reduce the magnetic disc disruption.
But although the central $dm/dr$ model in Figure \ref{fig:dmdrvary} has a factor ten higher wind mass loss rate, and thus a factor ten lower $\eta_\ast$, it nonetheless shows a disc disruption that is still quite similar\footnote{We have also done additional simulations with lower $\eta_\ast$, even below unity, and again find that the wind parameters have little effect on the disc disruption.} to the standard model case with a lower ${\dot M}$.
It thus seems, again somewhat surprisingly, that the radiatively driven wind plays at most a relatively a minor role in the magnetic disruption of the disc from such a B2e star.

\subsubsection{Effect of split monopole field topology}
\label{sec:splitmon}

The second panel from the right of Figure \ref{fig:dmdrvary} shows a similarly weak effect for a model that actually imposes such a split-monopole field topology right from the stellar surface, independent of any effects of the stellar wind.
Again, because this effectively eliminates the magnetic flux through the equatorial disc, one would expect this to significantly reduce the magnetic disc disruption.
Indeed, the initial evolution for this case does not show the strong initial ejection stream associated with the poloidal field threading the disc; but over the long term, the overall level of disc disruption is similar to what's seen in the standard case with a dipole field.
Again, somewhat surprisingly, a monopole vs. dipole field topology does not seem to have much effect on the magnetic disruption of the disc.

\subsubsection{Effect of reduced grid resolution, and associated enhancement in numerical reconnection}
\label{sec:gridres}

The rightmost panel of Figure \ref{fig:dmdrvary} shows a similarly weak effect for a simulation with a factor two reduction in grid resolution (with now only 50 latitudinal zones from pole to pole, but still with a relatively high concentration about the equator). Because this increases the effective scale for numerical reconnection of field lines, and so reduces the effective magnetic Reynolds number, one might expect this would lead to a somewhat weaker coupling between the field and gas, and so perhaps a reduced level of magnetic disc disruption.
But the results for this case in the rightmost panel are again very similar to the leftmost model with standard grid resolution.
Within the tested range, the enhanced magnetic dissipation and reconnection in the low-resolution model thus does not seem to have much effect on the disc disruption.

\section{Discussion, Conclusions and Future Work}

A key result of the MHD simulations here is that, for typical parameters for a Be star and disc, a modest  polar surface field strength of just $10$\,G is sufficient to substantially disrupt the inner disc, while fields of about $100$\,G, comparable to upper limits from spectropolarimetric observations, should effectively destroy the disc over about 1000\,ks, or a couple dozen near-star orbital periods $P_{\rm orb} \approx 40$\,ks\,$\approx 0.5$\,d.
Since such fields are thus inherently incompatible with the Keplerian, circumstellar discs that give rise to the signature Balmer emission of Be stars, this provides an attractive explanation for why Be stars are essentially never observed to harbor the large-scale fields that are inferred in 8-10\% of other, normal OBA stars on the main sequence.

Our parameter study shows, moreover, that the efficacy of magnetic destruction of such Keplerian discs depends mainly on just the characteristic plasma beta, and is largely insensitive to the variations of other parameters, including stellar rotation speed, wind mass loss rate, monopole vs.\ dipole field geometry, and reduced grid resolution that augments numerical reconnection.
When characterized in terms of the equatorial surface field strength and peak disc density near the stellar surface in the equatorial mid-plane, significant disc disruption can occur for moderately large $\beta_{\rm o} \approx 20$, with complete destruction for $\beta_{\rm o}$ of order unity.

Because Be discs are thought to form from {\em decretion} of material from the near-critically rotating star, any magnetic field that is sufficiently strong to dominate the near-star disc dynamics seems also likely to inhibit the formation of a Keplerian disc by mass ejections and/or eruptions; but further dynamical simulations accounting directly for such mass ejections are needed to confirm this.

The simulations here do not, however, in any way preclude the formation of {\em accretion} discs around stars with fields that are strong enough to dominate the near-star dynamics; the strong radial decline of field energy ($B^2 \sim r^{-6}$ for a dipole field)
means that such external, large-scale fields  have negligible effect on the outer disc, and thus on the source regions of the accreting disc mass.

But even for such accretion discs, which in principle can have much larger densities, the study here can provide an overall characterization of the field strengths needed to disrupt or erode the inner disc, and thereby perhaps compete with the accretion feeding rate from the outer disc.
They thus provide a simple but useful complement to other, more complete simulations of the effect of magnetic fields on the accretion process.

Further work is needed to relax the several simplifying assumptions of the present simulations.
Since the disc temperature affects its gas pressure and thus the plasma beta, future simulations should relax the assumption of isothermal plasma to include a realistic energy equation that accounts for the balance between radiative heating from the star and cooling emission from the disc, including a potentially optically thick radiation transport.
For the near critical rotation of Be stars, the assumption of uniformly bright spherical star should be relaxed to account for the effects of oblateness and gravity darkening.
Beyond the present 2D axisymmetric geometry for a rotation-aligned dipole field, future simulations should explore the effect of tilted dipoles, and/or more complex, higher-order multipole fields, that are usually more difficult to detect.
Each of these generalization could plausibly alter the nature and efficiency of magnetic disruption of the circumstellar disc, although the relative insensitivity found here for parameter variations beyond the characteristic plasma beta suggests that such generalized models might likewise follow similar overall scaling with field strength through a characteristic disc plasma beta.

Within these caveats, the central conclusion from this study remains that relatively modest large-scale magnetic field strengths near and below the current upper observational limits for Be stars seem likely to disrupt or destroy any circumstellar disc.
This thus provides a fundamental rationale for why Be stars remain the only class of early-type stars without any such observationally inferred large-scale, organized magnetic fields.

\section*{Acknowledgements}

A.uD acknowledges support by NASA through Chandra Award numbers GO5-16005X and TM7-18001X issued by the Chandra X-ray Observatory Center which is operated by the Smithsonian Astrophysical Observatory for and on behalf of NASA under contract NAS8- 03060.

\end{document}